\documentclass[aip,jmp,unsortedaddress,
preprint
]{revtex4-1}

\usepackage[english]{babel}
\usepackage[reqno]{amsmath}

\newcommand{\jcs}[1]{\mbox{$\tilde{J}_{#1}(x,y) $} }
\newcommand{\jbar}[1]{\mbox{$\widehat{J}_{#1}(y_1,y_2) $} }
\newcommand{\e}{\mbox{$\mathrm{e} $} }
\renewcommand{\d}[1]{\mbox{$\delta_{n,#1}$} }
\newcommand{\somma}[1]{\sum_{#1=-\infty}^{\infty}}
\newcommand{\summa}[1]{\sum_{#1=-\infty}^{\infty}}

\begin{document}

\title{A new class of sum rules for products of Bessel functions }

\author{G. Bevilacqua }
\email{bevilacqua@unisi.it}
\author{V. Biancalana }
\author{Y. Dancheva }
\affiliation{CNISM and  Dipartimento di Fisica, Universit\`a di Siena,
  Via Roma 56, 53100 Siena, Italy}

\author{ T. Mansour }
\email{toufik@math.haifa.ac.il}
\affiliation{ Department of Mathematics, University of Haifa, 31905 Haifa, Israel}

\author{L. Moi } 
\affiliation{CNISM and  Dipartimento di Fisica, Universit\`a di Siena,
  Via Roma 56, 53100 Siena, Italy}

\begin{abstract} 
  We derive a new class of sum rules for products of 
  Bessel functions of the first kind. Using standard algebraic
  manipulations we extend some of the well known properties of
  $J_n$. Some physical applications of the results are also
  discussed. A comparison with the
  Newberger[J. Math. Phys. \textbf{23} (1982) 1278] sum rules is
  performed on a typical example.
\end{abstract}

\date{\today}

\maketitle

\section{Introduction}
\label{sec:intro}

Bessel functions of the first kind ($J_n$) are  among the most ubiquitous
special functions of  mathematical physics. Their properties are
described in several
monographs\cite{Gray1895,Watson1922,Korenev2002,Dattoli1996}, in the
encyclopedic Abramowitz \& Stegun\cite{abramowitz:stegun} as well as
in every advanced mathematics textbook.

The generating
function identity  
\begin{equation}
  \label{eq:gen:fun}
\e^{\frac{z}{2}\left(w - \frac{1}{w}\right)} \equiv 
\sum_{n=-\infty}^{\infty} J_n(z) \;w^n
\end{equation}
is the key formula used in discussing the frequency modulation (FM) of
the electric field of a laser.  
In fact, a pure sinusoidal FM reads as  

\begin{equation}
  \label{eq:laser:mod}
  \mathbf{E}(t) = \frac{\mathbf{E}_0}{2}\left( \e^{i\left(\omega_L \,
        t + M \sin(\Omega \, t) \right)} + c.c. \right),
\end{equation}
where $\omega_L$ is the laser carrier frequency, $\Omega$ is the
modulation frequency, $M$ the modulation index and $M \Omega$ the
modulation depth. Using (\ref{eq:gen:fun}) with $w=\e^{i \,\Omega \, t}$
one finds the well-known sidebands structure
\begin{equation}
  \label{eq:sidebands}
   \mathbf{E}(t) = \frac{\mathbf{E}_0}{2}  \sum_{n=-\infty}^{\infty}
   J_n(M) \; \e^{i (\omega_L  + n \Omega) t } + c.c.
\end{equation}
In some applications, for instance studying the response of a
two-level system to a FM field, one has to deal with expressions like
\begin{equation}
  \label{eq:segnale}
  A_s = \somma{n} \frac{J_n(M) J_{n-s}(M)}{\gamma + i\, n\, \Omega} ,
\end{equation}
where $\gamma$ represents a resonance
linewidth. 
Even if the sum could be done exactly using the Newberger sum
rule\cite{newberger}, the result is quite involved 
(the details are given in Section~\ref{sect:newberger}). 

Alternatively, in the physical interesting limit $\epsilon= M
\Omega/\gamma < 1 $ 
(see below for more details), a geometric expansion
gives
\begin{equation}
  \label{eq:expa:geom}
  A_s = \frac{1}{\gamma} \somma{n} 
    J_n(M) J_{n-s}(M)
    \left[ 1 - \frac{i \Omega}{\gamma} n
      -  \left(\frac{\Omega}{\gamma}\right)^2 n^2 + 
      O(\epsilon^3)
    \right]
\end{equation}
reducing the problem to the evaluation of sums of the form
\begin{equation}
  \label{eq:sum:type}
  B_{k,s} \equiv \sum_{n=-\infty}^{+\infty} n^k J_n(M) J_{n-s}(M)
\end{equation}
for integer values of $k \geq 0$ and $s$. To the authors' best knowledge these
sums are not addressed in the literature in the general case although special
values are known, for instance $B_{0,0} = 1$ from 9.1.76 of
[\onlinecite{abramowitz:stegun}]. 

Finding a closed expression for $B_{k,s}$ is the main goal of the
present work.  The paper is organized as follows. In
Section~\ref{sec:res} the main results are derived and discussed,
while in Section~\ref{sec:appl} some applications are pointed
out. Finally Section~\ref{sec:concl} contains the conclusions. 

\section{Results}
\label{sec:res}

We start the derivation of the main result by rewriting
(\ref{eq:gen:fun})
in a slightly different form 
\begin{equation}
  \label{eq:gen:sen}
  f(y,\theta) \equiv \e^{i\, y\, \sin \theta} =
  \sum_{n=-\infty}^{\infty} J_n(y)\, \e^{i\, n\, \theta}.
\end{equation}
Deriving the left-hand side  $k$ times, one gets
\begin{equation}
  \label{eq:der:k}
  f^{(k)}(y,\theta) \equiv 
\frac{\partial^k f(y,\theta)}{\partial \theta^k} = p_k(y,\theta) f(y,\theta)
\end{equation}
where $p_k(y,\theta)$ is a polynomial in $\cos \theta$ and $\sin
\theta$ and can be easily recast in the 
form 
\begin{equation}
  \label{eq:form:pk}
  p_k(y,\theta) = \somma{n} C_{k,n}(y) \, \e^{i\, n\, \theta},
\end{equation}
with $C_{k,n}(y)$ polynomials in $y$. An explicit form for these
coefficients is worked out in Appendix~\ref{sec:coeff:Ckn}. 
Combining everything  one finds  
\begin{equation}
  \label{eq:fk:Jn}
    f^{(k)}(y,\theta) =  \somma{n} 
\left( 
  \somma{q} C_{k,q}(y) \, J_{n-q}(y) 
\right) \,
\e^{i\, n \, \theta}
\end{equation}

On the other hand, deriving the right-hand side of (\ref{eq:gen:sen})
one obtains 
\begin{equation}
  \label{eq:der:sum}
   f^{(k)}(y,\theta) = \sum_{n=-\infty}^{\infty} (i \, n)^k \, J_n(y)\, \e^{i\, n\, \theta}.
\end{equation}
An explicit comparison of (\ref{eq:der:sum}) with (\ref{eq:fk:Jn}) is
performed in Appendix~\ref{sec:rec:rel}. 
Now consider the expression   
$F(y_1, y_2, \theta)  \equiv f^{(k)}(y_1, \theta)  f(y_2,
\theta)$. A first form can be obtained 
using (\ref{eq:form:pk}) and (\ref{eq:gen:sen}) as follows  
\begin{equation}
  \label{eq:def:F}
  \begin{split}
    F(y_1, y_2, \theta) &=  p_k(y_1, \theta) \, \e^{ i \, (y_1 + y_2)\, \sin \theta } \\
    &= \left( \summa{n} C_{k,n}(y_1) \e^{i\, n\, \theta} \right)
    \left( \summa{m} J_m(y_1 + y_2) \e^{i\, m\, \theta} \right) \\
    &= \summa{q} \left( \summa{m} C_{k,q-m}(y_1) \, J_{m}(y_1+y_2) \right)
    \e^{i \, q \, \theta}.
  \end{split}
\end{equation}
In the meantime an equivalent formula can be written using
(\ref{eq:der:sum})
\begin{equation}
  \label{eq:def:F:Jn}
  \begin{split}
    F(y_1, y_2, \theta) 
    &= \left( \summa{n} (i\, n)^k \, J_n(y_1)\, \e^{i\, n\, \theta} \right)
    \left( \summa{m} J_m( y_2) \e^{i\, m\, \theta} \right) \\
    &= \summa{q} \left( \summa{n} (i \, n)^k \, J_{n}(y_1)\, J_{q-n}(y_2) \right)
    \e^{i \, q \, \theta},
\end{split}
\end{equation}
and, given the completeness of the set 
$\{ \e^{i\, q \, \theta}, \; q=0,\pm 1, \pm 2, \ldots  \}$, the 
coefficients in both expansions must be equal
\begin{equation}
  \label{eq:res:fond}
   \summa{m} C_{k,q-m}(y_1) \, J_{m}(y_1+y_2) =  \summa{n} (i \, n)^k \,
   J_{n}(y_1)\, J_{q-n}(y_2). 
\end{equation}
This is the central result of the paper and can be considered to be a
generalization of the $J_n$ addition formula
[\onlinecite[formula 9.1.75]{abramowitz:stegun}], which, in fact, is obtained for
$k=0$ noticing that $C_{0,n} = \d{0}$ (see Appendix \ref{sec:coeff:Ckn}). 
Besides, with a proper choice of $y_1$
and $y_2$, different sum rules can be obtained.  In fact, 
substituting  $y_1 = -y_2 = y$ in (\ref{eq:res:fond}) and remembering that 
$J_n(0) = \d{0}$ one finds  
\begin{equation}
  \label{eq:res:fond:1}
  \begin{split}
    C_{k,q}(y) & =  \summa{n}  (i\,n)^k \, J_{n}(y)\, J_{q-n}(-y) \\
    & =   \summa{n}  (i\,n)^k \,    J_{n}(y)\, (-1)^{q-n}\,J_{q-n}(y) \\
    & = \summa{n}  (i\,n)^k \,    J_{n}(y)\,J_{n-q}(y).
    \end{split}
\end{equation}
This gives an answer for the initial problem, namely
\begin{equation}
  \label{eq:Bks:soluz}
  B_{k,s} = \frac{1}{i^k}\, C_{k,s}
\end{equation}
and the  $B_{k,s}$  for the first $k$ values 
can be read from Appendix~\ref{sec:coeff:Ckn}. 

The companion case $y_1 = y_2 = y$ is worked out with similar 
algebraic elaborations leading to the formula
\begin{equation}
  \label{eq:sum:2y}
  \summa{n} \, (-1)^n \, n^k \, J_n(y)\, J_{n-q}(y) 
  = \frac{(-1)^q}{i^k} \, \summa{m} C_{k,q-m}(y) \, J_m(2y).
\end{equation}

\subsection{Generalizations}
\label{sec:generalizations}
Many generalizations are possible once one dives into the world of
generalized Bessel functions\cite{Dattoli1996}. Let us discuss some 
cases.

Consider the function
\begin{equation}
  \label{eq:mod:sincos}
  \e^{i\,(x\,\cos \theta + y\,\sin\theta)} \equiv 
\summa{n} \,  \jcs{n}\, \e^{i\,n\,\theta},
\end{equation}
which arises in the context of general sinusoidal modulation. The
$\jcs{n}$ are a kind of ``generalized'' Bessel functions, which can be
given in terms of the usual Bessel functions by means of simple algebraic
manipulations (see also equation (11.10) in Dattoli\cite{Dattoli1996})
\begin{equation}
  \label{eq:jcs:toJ}
  \jcs{n} \equiv \summa{q} \, i^q \, J_q(x)\, J_{n-q}(y). 
\end{equation}

The same steps as before lead to the sum rule
\begin{equation}
  \label{eq:jcs:sum:rule}
 2 \summa{n} n \, \jcs{n} \left[\jcs{n-q}\right]^* \equiv  
(y + i\,x) \, \delta_{q,1} + (y - i\,x) \, \delta_{q,-1} .
\end{equation}
Furthermore, $k$-order derivatives lead to sum rules with $n^k$ which the
interested reader can easily work out. 

The second generalization is achieved considering 
\begin{equation}
  \label{eq:mod:sinsin}
  \e^{i\,(y_1\,\sin \theta + y_2\,\sin 2\theta)} \equiv 
\summa{n} \,  \jbar{n}\, \e^{i\,n\,\theta}.
\end{equation}
These $\jbar{n}$ ``generalized'' Bessel functions can be expressed as 
\begin{equation}
  \label{eq:jbar:toJ}
  \jbar{n} \equiv \summa{q}  \, J_q(y_2)\, J_{n-2\, q}(y_1), 
\end{equation}
and are real functions for real values of $y_i$ 
(see chapter 2 of Dattoli\cite{Dattoli1996} for further details). 
By means of the same device one gets
\begin{equation}
  \label{eq:jbar:sum:rule}
  \summa{n} n \, \jbar{n} \jbar{n-s} \equiv  
\frac{y_1}{2} (\delta_{s,1} + \delta_{s,-1}) +
y_2  (\delta_{s,2} + \delta_{s,-2}).
\end{equation}
Also in this case higher derivatives lead to other sum rules. 

In a complete general setting, 
one could consider the problem of a 
FM signal with  no special properties, i.e. 
\begin{equation}
  \label{eq:gen:mod:sign:exp}
  \e^{i\, \phi(t)} = \summa{n} G_n \, \e^{i\,n\,\Omega\, t}
\end{equation}
where the only requirement is that $\phi(t)$ is a real and 
periodic well-behaved function
(a more precise statement is given below)
\begin{equation}
  \label{eq:gen:mod:sign}
  \phi(t) = \summa{n} \, \phi_n\, \e^{i\,n\,\Omega\,t}.
\end{equation}
The $G_n$ coefficients are related in a very complicated way to multiple
sums of products of ordinary Bessel functions, but the explicit form
is not important here. They satisfy the ``conservation of energy'' sum
rule
\begin{equation}
  \label{eq:gen:mod:cons:ene}
  \summa{n} G_n \, G_{n-s}^* = \delta_{s,0}.
\end{equation}

Taking the time derivative of (\ref{eq:gen:mod:sign:exp}) and doing the
same steps as above we get
\begin{equation}
  \label{eq:gen:mod:sum:rule}
  \summa{n} n\, G_n \, G_{n-s}^* = i\, s\, \phi_s.
\end{equation}
Equation~(\ref{eq:gen:mod:sum:rule}) holds provided that
$\dot{\phi}(t)$ can be expanded as a meaningful Fourier series, i.e. 
\begin{equation}
  \label{eq:gen:mod:sign:der}
  \dot{\phi}(t) = \summa{n} \, i\,n\,\Omega\,\phi_n\, \e^{i\,n\,\Omega\,t}
\end{equation}
must be convergent. 

As before, higher order derivatives will give higher order sum rules.

\section{Applications}
\label{sec:appl}
An  example is considered where formula (\ref{eq:Bks:soluz}) comes in 
handy. Let us model an optical resonance as a damped harmonic
oscillator of unit mass,
forced by the electric field of the light beam
\begin{equation}
  \label{eq:appli:ini}
  \ddot{z} + \gamma \dot{z} + \omega_0^2 z = f \e^{+i \omega t},
\end{equation}
where, as usual, the physical oscillator 
displacement  is $x(t) = \Re
(z(t))$ and all other symbols have a clear meaning. 
Such a simple and analytical model avoids all the complications
tied to the solution and lets us concentrate on the sum rule. In
fact, after the transient, the solution settles into the  well-known
\begin{equation}
  \label{eq:appli:sol:z}
  z(t) = \frac{f\, \e^{+i \omega t}}{\omega_0^2 - \omega^2 + i \gamma \omega } 
,
\end{equation}
and the work done by the light field per time unit is 
\begin{equation}
  \label{eq:appli:ass}
  \frac{d W}{dt}  = \frac{f^2}{4} 
  ( \e^{i \omega t} + \e^{-i \omega t})
  \left(
    \frac{i \omega \, \e^{i \omega t}}{\omega_0^2 - \omega^2 + i \gamma \omega}
     +
    \frac{-i \omega \, \e^{-i \omega t}}{\omega_0^2 - \omega^2 - i \gamma \omega}
  \right).
\end{equation}
The experimental apparatus usually averages out high frequencies and
thus the measured signal is
\begin{equation}
  \label{eq:appl:ass:media}
  \langle  \frac{d W}{dt} \rangle = 
-  \frac{f^2}{2} \Im
\left( \frac{\omega}{\omega_0^2 - \omega^2 + i \gamma \omega} 
\right) 
=  \frac{f^2}{2} \frac{\omega^2 \gamma}{(\omega^2-\omega_0^2)^2 +
  \omega^2 \gamma^2}, 
\end{equation}
which, for nearly resonant light $\omega = \omega_0 + \delta$, $ \delta
\ll \omega_0$, becomes
\begin{equation}
  \label{eq:appl:ass:svil:lore}
  \langle  \frac{d W}{dt} \rangle 
  = \frac{f^2}{2} \left[ 
\frac{1}{\gamma}\,\frac{1}{1+\Delta^2} + 
\frac{1}{2}\,\frac{\Delta^3}{\left(1+\Delta^2\right)^2} \, \frac{1}{\omega_0} 
+ O\left(\frac{1}{\omega_0^2}\right) \right],
\end{equation}
where $\Delta=2\delta/\gamma$. It is easily seen that the leading term  
shows the well-known Lorentzian form.  

Replacing the forcing term with a frequency-modulated one
\begin{equation}
  \label{eq:appl:mod}
  \e^{i\,\omega\,t } \rightarrow   
\e^{i\,\left(\omega\,t +M \sin \Omega t\right)}
 = \sum_{n=-\infty}^{+\infty} \, J_n(M) \, \e^{i\, \omega_n\, t}
\end{equation}
on the right-hand side of (\ref{eq:appli:ini}) we get the exact result
\begin{equation}
  \label{eq:appl:ass:mod}
  \langle  \frac{d W}{dt} \rangle = 
- \frac{f^2}{2} \, \Im \left[ 
  \summa{s} 
  \left( 
    \summa{n} \, 
    \frac{\omega_n\, J_n(M) \, J_{n-s}(M)}{\omega_0^2 - \omega_n^2 + i\gamma\omega_n} 
  \right)
  \e^{i\, s \, \Omega\, t}
\right]
\end{equation}
where $\omega_n=\omega_0 + \delta +n\,\Omega$. To proceed any further
it is necessary to simplify the expression $F =   \omega_n/( \omega_0^2 - \omega_n^2 +
i\gamma\omega_n) $. 
In the optical range it is not restrictive to assume
$\omega_0 \gg \gamma, \delta $ thus getting
\begin{equation}
  \label{eq:svil:F}
  F = -\frac{1}{2\delta+2\,n\Omega-i\,\gamma}\,
  \left[ 1 + \frac{(\delta+n\Omega)^2}{2\delta+2\,n\Omega-i\,\gamma}\,
    \frac{1}{\omega_0} +
O\left( \frac{1}{\omega_0^2} \right) \right].
\end{equation}
Considering only the leading term in $F$ we are faced in
(\ref{eq:appl:ass:mod})
 with the sums
\begin{equation}
  \label{eq:ass:F:svil}
  \frac{1}{i\gamma}   
  \summa{n} \frac{J_n(M) J_{n-s}(M)}{1+i\Delta+ 2\, n\, i\, \Omega/\gamma}
\end{equation}
which are very similar to (\ref{eq:segnale}). To proceed any further
with the geometric expansion of the denominator, we notice that the
product  $J_n(M)\,J_{n-s}(M)$ shows the well-known ``bridge''
structure  and it is exponentially small when 
$|n| > N_{MAX} \approx  2\,M$. So if 
\begin{equation}
  \label{eq:svil:denom}
|2\, N_{MAX} \Omega /\gamma| < 1  
\;\; \Rightarrow
\left| \frac{2\, N_{MAX} \Omega /\gamma}{1+ i\,\Delta} \right|  < 1
\end{equation}
and the sum can be elaborated as follows
\begin{equation}
  \label{eq:sum:elab}
\begin{split}
  \frac{1}{i\gamma}   
  \sum_{n=-\infty}^{\infty}
  \frac{J_n(M) J_{n-s}(M)}{1+i\Delta+ 2\, n\, i\, \Omega/\gamma}  &
  \approx
  \frac{1}{i\gamma}   
  \sum_{n=-N_{MAX}}^{N_{MAX}}
  \frac{J_n(M) J_{n-s}(M)}{1+i\Delta+ 2\, n\, i\, \Omega/\gamma}
  \\
  & = \frac{1}{i\gamma}   
  \sum_{n=-N_{MAX}}^{N_{MAX}}
  \frac{J_n(M) J_{n-s}(M)}{1 + i\,\Delta} \,
  \left[ 1 +  n \epsilon + n^2 \epsilon^2  + O(\epsilon^3)
    \right] \\
 &  \approx   \frac{1}{i\gamma}   
      \sum_{n=-\infty}^{\infty}
      \frac{J_n(M) J_{n-s}(M)}{1 + i\,\Delta} \,
      \left[ 1 +  n \epsilon + n^2 \epsilon^2  + O(\epsilon^3)
      \right],
\end{split}
\end{equation}
where $\epsilon=- i \,(2 \Omega/\gamma)/(1+i\Delta)$ and the last row
is justified because of the exponentially small nature of the terms
included. 

The inner sum in (\ref{eq:appl:ass:mod}) can now worked out by means of
the sum rules developed above, and, after some straightforward
algebra, one finds
\begin{equation}
  \label{eq:appl:mod:fin}
\begin{split}
  \langle  \frac{d W}{dt} \rangle =  
  \frac{f^2}{2} \, \frac{1}{\gamma} &\left\{ 
    \frac{1}{1+\Delta^2}  +\frac{2 M \Omega}{\gamma}\,
    \frac{-2\Delta}{(1+\Delta^2)^2}\,
    \cos \Omega  t \right. \\
    & \;\;\; +\frac{1}{M} \left(\frac{2 M \Omega}{\gamma}\right)^2
    \frac{\Delta (\Delta^2 -3)}{(1+\Delta^2)^3}
    \, \sin \Omega t\\
    & \left. \;\;\; +\frac{1}{2} \left(\frac{2 M \Omega}{\gamma}\right)^2
      \frac{ 3\Delta^2-1}{(1+\Delta^2)^3}\,
      \, (1+\cos 2 \Omega t) + \ldots  \right\},
\end{split}
\end{equation}
where the terms in parentheses, besides the d.c. component, represent
the higher harmonic modulated absorption. Notice how the $\Delta$
dependence of the $(2M\Omega/\gamma)^n$ term is related to the real
and imaginary part of $n$-th derivative of $1/(1+i\Delta)$. It seems
worth stressing that the validity of (\ref{eq:appl:mod:gen:fin}) is
limited to modulation depths smaller than the linewidth as pointed out
in (\ref{eq:svil:denom}). Usually in typical experimental conditions 
 one is interested in the first harmonic signal,
catching it, for instance, by means of a lock-in amplifier. The above
formula shows that it is more convenient to use the  biggest value of
the  modulation depth $M\, \Omega$ compatible with (\ref{eq:svil:denom}). 

More generally, using an arbitrary but limited  modulation signal as in
(\ref{eq:gen:mod:sign}) one finds
\begin{equation}
  \label{eq:appl:mod:gen:fin}
  \begin{split}
    \langle  \frac{d W}{dt} \rangle & =  
     \frac{f^2}{2} \, \frac{1}{\gamma} \left\{   
      \frac{1}{1+\Delta^2}  
      +\frac{2}{\gamma} \frac{-2\Delta}{(1+\Delta^2)^2}\,  \dot{\phi}(t)
      + \ldots \right\} \\
    & =  
     \frac{f^2}{2} \, \frac{1}{\gamma} \left\{   
      \frac{1}{1+\Delta^2}  
      +\frac{2 \sigma }{\gamma}
      \frac{-2\Delta}{(1+\Delta^2)^2}\,  
      \frac{\omega(t)}{\sigma}
      + \ldots \right\}.
  \end{split}
\end{equation}
where in the last row the instantaneous frequency 
$\omega(t) = \dot{\phi}(t)$
 is introduced and the ``modulation depth'' 
$\sigma=|\omega_{MAX} - \omega_{min}|$ is put in evidence to compare
with previous formulae.


\subsection{Comparison with the Newberger sum rule}
\label{sect:newberger}

As stated in the Introduction, the quantity $A_s$ can be summed by
means of the Newberger\cite{newberger} sum rule. The result is
\begin{equation}
  \label{eq:newb:sum:As}
  A_s = \frac{(-1)^{s}}{\gamma} 
  \frac{\pi\,\gamma/\Omega}{\sinh(\pi\,\gamma/\Omega)}\,
   J_{s-i\,\gamma/\Omega}(M)\,J_{i\,\gamma/\Omega}(M),
\qquad s\ge 0
\end{equation}
where one has to deal with complex-order Bessel functions. Even if
this is an exact result, the physics is obscured by the complexity of
the formula. For instance, it is not easy to extract the behaviour when
$\eta=\Omega/\gamma$ is small. 
Instead using (\ref{eq:expa:geom}) with $s=1$ we quickly found
\begin{equation}
  \label{eq:appl:compar}
A_1 = -\frac{M}{2\,\gamma}\, i\, \eta  -
\frac{M}{2\,\gamma}\,\eta^2 +
\frac{M}{2\,\gamma}\, (1+3\,M^2/4)\, i\, \eta^3 +
O(\eta^4 )
\end{equation}

To proceed further using (\ref{eq:newb:sum:As})  the
product of the Bessel
functions must be developed\cite[formula 9.1.14]{abramowitz:stegun},
and, making use of the properties of the Euler 
Gamma function\cite[formula 6.1.31]{abramowitz:stegun},  one gets
\begin{equation}
  \label{eq:newb:gamma:res}
  A_s = \frac{(-1)^s}{\gamma}\, \left(\frac{M}{2}\right)^s
  \sum_{k=0}^{\infty} \frac{ (-\frac{M^2}{4})^k (s+2k)!}{(s+k)!\,k!}  \\
  \left( \prod_{p=1}^s \frac{1}{ k+p-i\,\gamma/\Omega }
  \prod_{p=1}^k \frac{1}{ p^2+\,(\gamma/\Omega)^2 } \right)
\end{equation}
which, to the authors' best knowledge, is an original elaboration. 
This formula is still complicated, but better suited to extracting
asymptotic behaviour. In fact for $s=1$ we have
\begin{equation}
  \label{eq:newb:A1}
\begin{split}
 A_1 & =-\frac{M}{2\gamma}
\left( 
\frac{1}{1-i\gamma/\Omega} -
\frac{3 M^2}{4}
\frac{1}{2-i\gamma/\Omega}\, \frac{1}{1+(\gamma/\Omega)^2}
+ \ldots
\right) \\ 
& =  -\frac{M}{2\,\gamma}\, i\, \eta  -
\frac{M}{2\,\gamma}\,\eta^2 + 
\frac{M}{2\,\gamma} (1 + 3\,M^2/4) \, i\, \eta^3 
+ O(\eta^4).
\end{split}
\end{equation}

\section{Conclusions} 
\label{sec:concl}

We have derived some new classes of sum rules and recursion relations
obeyed by  Bessel functions of the first kind. The results have broad
applications in the physics of modulated excitation for instance in the
case of light-matter interaction. 
A comparison of our results with an elaborated form of the
Newberger sum rule is also given.

\appendix

\section{Determination of $C_{k,n}$}
\label{sec:coeff:Ckn}

The starting point is the recursion obeyed by the $p_k$
polynomials. In fact, it is easily seen that $p_0 \equiv 1$, and
assuming (\ref{eq:der:k}) then follows
\begin{equation}
  \label{eq:der:kp1}
  p_{k+1}(y,\theta) = \left( 
    \frac{\partial}{\partial \theta} + i\, y\, \cos \theta
    \right) \, p_k(y, \theta). 
\end{equation}
Projecting this equation as shown in (\ref{eq:form:pk}) one obtains
\begin{equation}
  \label{eq:coeff:Ckn:eqz}
  C_{k+1,n} = i\,n \,C_{k,n} + i\,\frac{y}{2}
    \left( 
      C_{k,n+1} + C_{k,n-1}
    \right)  
\qquad \mathrm{with} \qquad
C_{0,n} = \delta_{n,0}.
\end{equation}
To the authors' best knowledge this recursion cannot be solved in terms
of known special functions, even if the homogeneous right-hand side,
namely $ i\,n \,C_{k,n} + i\,\frac{y}{2} \left( C_{k,n+1} + C_{k,n-1}
\right) = 0 $, is satisfied by the $J_n(y)$ itself. 
The first few loops give
\begin{eqnarray*}
  \label{eq:Ckn:espl}
  \frac{1}{i} C_{1,n}  & = &\frac{y}{2} \left(  \d{+1} + \d{-1} \right) \\
  \frac{1}{i^2} C_{2,n}  & =& \frac{y^2}{4}\left( 2\d{0} + \d{+2} + \d{-2} \right)
  + \frac{y}{2}\left( \d{+1} - \d{-1} \right) \\
  \frac{1}{i^3}C_{3,n}  & =& \frac{y^3}{8}\left( \d{+3} +\d{-3} \right) + 
  \frac{3\,y^2}{4} \left( \d{+2} - \d{-2} \right) + \\ \nonumber
  &\phantom{=} & 
  \left( \frac{3\,y^3}{8} + \frac{y}{2} \right) \left( \d{+1} +
    \d{-1} \right) \\
  \frac{1}{i^4}C_{4,n} & = &\frac{y^4}{16}\left( \d{+4} + \d{-4} \right) + 
  \frac{3 y^3}{4} \left( \d{+3} - \d{-3} \right) + \\ \nonumber
  & \phantom{=} &\left( \frac{y^4}{4} +\frac{7y^2}{4} \right) \left( \d{+2} +
    \d{-2} \right) + \\ 
 & \phantom{=}  &   \left( \frac{3y^3}{4} +\frac{y}{2} \right) \left( \d{+1} -
    \d{-1} \right) + \\ \nonumber 
  & \phantom{=}  &\left( \frac{3y^4}{8} +\frac{y^2}{4} \right) \d{0}
\end{eqnarray*}
The recursion is straightforward, but tedious.  A simplification is
achieved thinking of (\ref{eq:coeff:Ckn:eqz}) as a matrix vector product
and implementing it in any computer algebra system.

Alternatively, an approach based on the Fa\`a di Bruno formula can be
developed. The Fa\`a di Bruno formula  generalizes the chain rule 
to higher derivatives and can be stated as follows
\begin{equation}
  \label{eq:FdB}
\begin{split}
&\frac{d^k}{d\,x^k} f(g(x)) \\
&=\sum_{ \{m_j\}} \frac{k!}{m_1!\,m_2!\,\cdots\,m_k!} f^{(m_1+\cdots+m_k)}(g(x)) \prod_{j=1}^k\left(\frac{g^{(j)}(x)}{j!}\right)^{m_j}.
\end{split}
\end{equation}
where the sum is over all $k$-tuples of non-negative integers 
$m_1,\ldots, m_k$ satisfying the constraint 
$ 1\, m_1+2\, m_2 + 3\,m_3 + \cdots +k\, m_k=k$.
 
Applying that device to our problem we find 
\begin{equation}
  \label{eq:pk:FdB}
\begin{split}
p_k & = \e^{-i\, y\, \sin \theta} \, \frac{d^k}{d\, \theta^k}\e^{i\, y\, \sin \theta} \\
 & = \sum_{ \{ m_j\}} \, \frac{k!}{\prod_j m_j! \,j!^{m_j} }\,(i\, y)^m \, \prod_{j=1}^k(\sin^{(j)}\theta)^{m_j},
\end{split}
\end{equation}
where $m= \sum_{j \geq 1} m_j$. Next using 
$\sin^{(2j-1)}\theta = (-1)^{(j-1)}\cos \theta$ and 
$\sin^{(2j)}\theta = (-1)^{(j)}\sin \theta$ we get 
\begin{equation}
  \label{eq:pk:1:FdB}
    p_k  = \sum_{ \{ m_j\}} \, \frac{k!}{\prod_j m_j! \,j!^{m_j} }\, 
(i\, y)^m \, (-1)^\phi \, \sin^a \theta \cos^b \theta
\end{equation}
where $\phi= \sum_{j \geq 0} ( m_{2 + 4\,j} + m_{2+4\,j +1} )$,
 $a = \sum_{j\geq 1} m_{2\,j}$ and 
$b = \sum_{j\geq 1} m_{2\,j - 1} = m - a$. Finally expressing the
trigonometric functions as 
\begin{equation}
  \label{eq:pk:FdB:trig}
  \sin^a \theta \, \cos^b \theta = \frac{1}{2^m \, i^a} 
  \sum_{r=0}^a \, \sum_{r'=0}^b \, (-1)^r  \binom{a}{r}
  \binom{b}{r'}
  \e^{i\,( m - 2 r - 2 r')\,\theta}
\end{equation}
and substituting in (\ref{eq:pk:1:FdB}) we find 
\begin{equation}
  \label{eq:Ckn:FdB}
    C_{k,n} = \sum_{ \{ m_j\} } \frac{k!i^b(-1)^\phi(y/2)^m}{ \prod_j m_j! \, j!^{m_j}} 
   \sum_{r=0}^{a} \, (-1)^r \,\binom{a}{r}
  \binom{b}{\frac{m-n}{2} -r}
\end{equation}
with $m\geq n\geq -m$, where we follow the convention that the
binomial coefficients are zero if the lower index is not an integer or
negative or larger than the upper one.

\section{Some new recursion relations}
\label{sec:rec:rel}

By comparing the two different forms (\ref{eq:fk:Jn}) and
(\ref{eq:der:sum}) of the same quantity, a relation between the
coefficients results
\begin{equation}
  \label{eq:ric:def}
  q^k \, J_q(y) \equiv \summa{n}\frac{1}{i^k} \, C_{k,n}(y) \,\, J_{q-n}(y).
\end{equation}
The $k=0$ case is the identity $J_q(y) = J_q(y)$. For $k=1$ we obtain
the well-known\cite[formula 9.1.27 first row]{abramowitz:stegun} three
term recursion relation  
\begin{equation}
  \label{eq:rec:k=1}
  2\, q\, J_q(y) = y \, \bigl[ J_{q+1}(y) + J_{q-1}(y) \bigr] . 
\end{equation}

The interesting and, to the authors' best knowledge, new relations are
found for $k\geq 2$. In fact for $k=2$ one gets
\begin{equation}
  \label{eq:rec:k=2}
\begin{split}
  &\left( \frac{y^2}{2} - q^2 \right)\, J_q(y) \\
  &=
\frac{y}{2} \bigl[  J_{q+1}(y) - J_{q-1}(y) \bigr] -
\frac{y^2}{4} \bigl[  J_{q+2}(y) + J_{q-2}(y) \bigr].
\end{split}
\end{equation}
For higher values of $k$, other relations are easily obtained using the
coefficients of Appendix \ref{sec:coeff:Ckn}.




%

\end{document}